# A FRAMEWORK FOR DETECTING FRAUDULENT ACTIVITIES IN EDO STATE TAX COLLECTION SYSTEM USING INVESTIGATIVE DATA MINING

Okoro F. M[1], Oshoiribhor E. O[2] & John-Otumu M. A[3]

[1]Department of Mathematics, Ambrose Alli University, Ekpoma, Nigeria [2]Department of Computer Science, Ambrose Alli University, Ekpoma, Nigeria [3]ICT Directorate, Ambrose Alli University, Ekpoma, Nigeria

*Abstract:*

*Edo State Inland Revenue Services is overwhelmed with gigabyte of disk capacity containing data about tax payers' in the state. The data stored on the database increases in size at an alarming rate. This has resulted in a data rich but information poor situation where there is a widening gap between the explosive growth of data and its types, and the ability to analyze and interpret it effectively; hence the need for a new generation of automated and intelligent tools and techniques known as investigative data mining, to look for patterns in data. These patterns can lead to new insights, competitive advantages for business, and tangible benefits for the State Revenue services. This research work focuses on designing effective fraud detection and deterring architecture using investigative data mining technique. The proposed system architecture is designed to reason using Artificial Neural Network and Machine learning algorithm in order to detect and deter fraudulent activities. We recommend that the architectural framework be developed using Object Oriented Programming and Agent Oriented Programming Languages.*

## 1. INTRODUCTION

The Edo State Board of Internal Revenue services was established by the Edo State Revenue Administration Law, 2012, effectively replacing Board of Internal Revenue which was established by the Personal Income Tax Act Cap.71 1972 Laws of Bendel State as amended in 1976 (applicable to Edo State). The Revenue Administration Law, 2012, also established an operational arm of the Board which is known as the Edo State Internal Revenue Service. The Board is statutorily charged with the following responsibilities:

(a) Collecting all taxes from public/private workers on their income.
(b) Collecting all levies and penalties accruable to the State Government from shops, stores, eateries, restaurants, licensing of motor vehicles, and so on.
(c) Advising on the structure, incidence and administration of taxes in accordance with the provision of the Personal Income Tax Act of 2004 (as amended) which is applicable all over the Federation.

The current manual method of collecting these taxes/levies from tax payers lack proper organization and prone to obvious fraudulent activities, thereby resulting to loss of funds by the State Government. This however, leads to the inability of the Government meeting the basic projects requirements like rehabilitation of roads, building/ renovation of hospitals, schools and so on. The manual method does not clearly states the strategy that is being used to apportion taxes or levies or charges to some key business enterprise like electronic shops, saloon, hawkers, petty shops, barbing saloon, recharge card resellers, cosmetics shops, super markets etc. Random amount is being collected as taxes/levies from these business enterprises, which might be below or over charged taxes; and most times the levies/taxes might not be remitted to the state Government or if remitted, a reasonable amount would have been diverted by the officers in charge of the collection. This situation occurs because there is no proper structure on ground to determine the exact monies and the number of petty shops or saloons in a given location, to enable the State Government to ascertain the actual figures, in terms of Naira and Kobo that is generated from such businesses in Edo state.

Fraud has become one of the constants of life; fraud can never be stopped but can rather be detected and reduced to an extent [1]. There is the need for a new generation of automated and intelligent tools and techniques[2] known as investigative data mining, to look for patterns in data that will group these taxpayers into various groups for the purpose of paying their correct tax amount as specified per group. These patterns can lead to new insights, competitive advantages for businesses and tangible benefits for the state Governments in terms of internally generated revenue.

During the course of analyzing the existing system of taxes/levies collection, the following problems were identified;

   i. Fraud activities were identified on monies collected from petty shops and other unorganized businesses in the state.
  ii. No proper designed structure for apportioning taxes/levies to petty shops, super markets saloons, and so on, in the existing system. That is, the tax intelligent officers give estimated levies or taxes to shop at random using their discretional power thereby under or over levying the taxpayers. For example, where there are two super markets A and B in the same location; Suppose, super market A is well stocked and sells more than super market B. The taxes for both super markets ought not to be the same. In the existing system the Tax intelligent officer apportions the same amount of tax to both super markets, which is not ideal. Tax is supposed to be based on a certain percentage of your earning or income or profit. In the scenario described above, super market A is supposed to pay more tax because more profit is being generated from super market A as compared to super market B.
 iii. Some of the monies collected from these petty shops as taxes/levies are not being remitted to the appropriate office, there by leading to internally generated revenue suppression or loss.

Though the existing system utilizes computers for some internal operations; the system and entire process is not still intelligent enough to detect or deter fraudulent activities. The computers installed are just used for storage of tax payers' information and the Government does not know the number of petty shops, offices and companies that are paying these specific taxes/levies; let alone the real revenue generated monthly from certain business ventures in certain location or areas. The State Government only makes decision based on what the Edo State board of internal revenue presented as revenue generated or collected for the month. It is important to note that the mode of operation in the proposed system architecture is using data mining technology to investigate, detect fraud activities and secure the systems data pool. The world is overwhelmed with millions of gigabyte disks containing data. It is estimated that these data stored in all corporate and government databases worldwide doubles every twenty months. The types of data available, in addition to the size, are also growing at an alarming rate. The data mining field spans several research areas [3] with stunning progress over the last decade. Database theories and tools provide the necessary infrastructure to store, access and manipulate data. Artificial intelligence research such as machine learning and neural networks are concerned with inferring models and extracting patterns from data. Data visualization examines methods to easily convey a summary and interpretation of the information gathered. Statistics is used to support and negate hypotheses on collected data and control the chances and risks that must be considered upon making generalizations. Distributed data mining deals with the problem of learning useful new information from large and inherently distributed databases where multiple models have to be combined. The most common goal of business data mining applications is to predict customer behavior.

However this can be easily tailored to meet the objective of detecting and preventing criminal activity. It is almost impossible for perpetrators to exist in this modern era without leaving behind a trail of digital transactions in databases and networks [4]. Therefore, investigative data mining is about systematically examining in detail, hundreds of possible data attributes from such diverse sources as law enforcement, industry, government, and private data provider databases. It is also about building upon the findings, results and solutions provided by the database, machine learning, neural networks, data visualization, statistics, and distributed data mining communities, to predict and deter illegitimate activity.

The specific objective of this paper is to critically examine some known fraud detection models in order to design a single framework for data capturing, cleansing, classification, detecting / deterring fraudulent activities, and reports for tax collection system in Edo State, Nigeria.

## 2. RELATED WORK

Literatures have shown that detecting clusters of crime incidents [5] and finding possible cause/effect relations with association rule mining [6] are important to criminal analysis. The classification techniques have also proven to be highly effective in fraud detection [7, 8] and can be used to categorize future crime data and to provide a better understanding of present crime data.

### 2.1 Some Existing Fraud Detection Methods

This section reviews some present data mining methods applied specifically to the data-rich areas of insurance, credit card, and telecommunications fraud detection. [9] Recommends the use of dynamic real-time Bayesian Belief Networks (BBNs), which was named Mass Detection Tool (MDT) for the early detection of potentially fraudulent claims. The weights of the BBN are refined by rule generator's outcomes and claim handlers which keep pace with evolving frauds. This approach evolved from ethnology studies of large insurance companies and loss adjustors who argued against the manual detection of fraud by claim handlers. The hot spot methodology according to [10] applied three step processes: the k-means algorithm for cluster detection, the C4.5 algorithm for decision tree rule induction, and domain knowledge, statistical summaries and visualization tools for rule evaluation. A similar methodology presented by [11] utilized the SOM for cluster detection before back propagation neural networks in automobile injury claims fraud.

According to [12] the distributed data mining model is another scalable, supervised black box approach that uses a realistic cost model to evaluate C4.5, CART, Ripper and naive Bayesian classification models. The results shows that partitioning a large data set into smaller subsets to generate classifiers using different algorithms, experimenting with fraud/legal distributions within training data and using stacking to combine multiple models significantly improves cost savings. This method was applied to one million credit card transactions from two major US banks, Chase Bank and First Union Bank.

The neural data mining approach by [13] uses generalized rule-based association rules to mine symbolic data and Radial Basis Function neural networks to mine analog data. Results shows that using supervised neural networks to check the results of association rules increases the predictive accuracy, while the credit fraud model by [14] recommends a classification approach when there is a fraud/legal attribute, or a clustering followed by a classification approach if there is no fraud/legal attribute. Similarly a review was also done on the application of artificial intelligence techniques in combating cyber-crimes [15].

However, research work conducted by [16] on the application of soft computing to tax fraud detection in small business using artificial neural network (ANN) technique as the detection mechanism. Suitable conclusions about the tax fraud were gotten based on market condition and periodical financial reports.

## 3. METHODOLOGY

This section explains the various methodologies used in realizing the designed framework for detecting fraudulent activities in Edo State Tax Collection System.

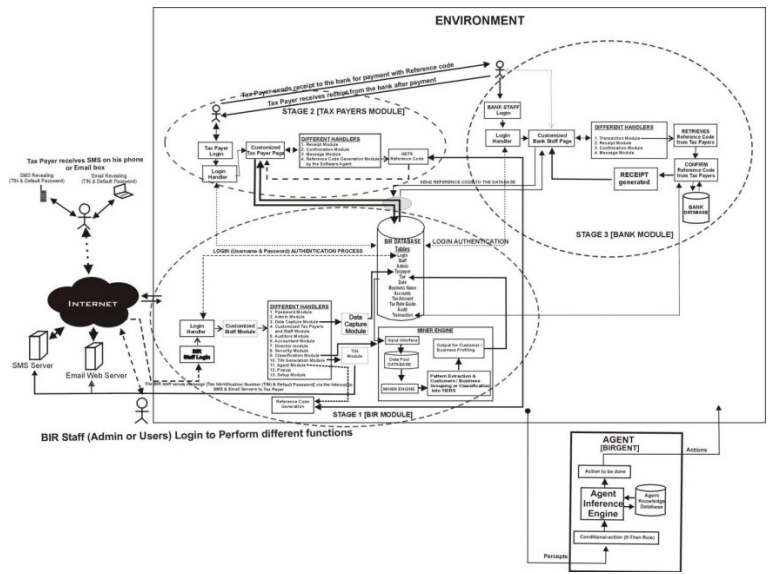

Figure 1. System Architecture for fraud control and detection in tax collection system

The diagram in Figure 1 depicts the proposed system architecture, which is sub divided into three main modules and a software agent called BIRGENT which is actually not seen, but relates with the system's environment / agency. The main functional modules for human interaction are (a) BIR (Board of Internal Revenue) Module, (b) Tax Payers' Module and (c) Bank Module. The intelligent software agent termed BIRGENT does not really relates with users in order to perform its function of fraud detection and prevention within the application. It is able to reason as designed using artificial neural network (ANN) and the engine is trained using supervised learning technique (profiling of known data to the knowledge base). The intelligent agent module interacts with the database directly in order spy on all data / transaction activities entries and analyzes them in order to detect and deter fraudulent activities like cash suppression or diversion by the generation of unique transaction reference code for tax payers with a well-structured tax payer's details information attached per transaction, which it also compares with the tax payer's TIN. The agent has well organized reasoning ability.

### 3.1 Complete Workflow Process for the 3 functional modules in the System Architecture

  i. BIR Module: Administrator creates staff / staff username and password in the system database.
 ii. BIR Module: BIR staff logon using the password and username given by the Administrator
iii. BIR staff capture all the taxpayer information into the data pool using captured page in the BIR Application
 iv. BIR staff classify taxpayers into tier using the Miner Engine.
  v. BIR Module: Administrator generates unique Tax Identification Number (TIN) against each taxpayer and a default password. The TIN and the password will be sent to the taxpayer email address provided or as a text message to their phone number
 vi. Taxpayer Module: Tax payer logon to the system using the TIN number and the default password.
vii. The system prompt the taxpayer to change the default password to his/her own personalize password.
viii. Tax payer clicks on pay tax from his/her personalized page and the taxpayer information will display including the tax amount.
 ix. Tax payer then click on Get Reference Code if he/she intends to pay tax. The software agent will generate the Reference Code.
  x. The unique Reference Code is displayed.
 xi. Print out the Reference Number including the tax amount, taxpayer name, Business name and date and take the print out to Bank to pay the tax amount
xii. Bank Module: Bank staff login using a username and password.
xiii. Bank Module: Bank staff type in the Reference code in the textbox provided in the module and click on search.
xiv. The taxpayer information is displayed which includes the name of the business, the taxpayer name, tax amount, TIN if the Reference code is genuine, and If detected by the software agent not to be genuine, no information will be displayed. If Reference code is stolen, the name of the original owner will be displayed.
 xv. Bank Staff collect cash equivalent from the taxpayer and click on paid in the BIR bank Module and a receipt will be printed out and given to the Taxpayer.
xvi. The receipt will consist of the business name, amount paid, date and the same Reference Code etc.

| | |
|---|---|
| xvii. | Bank staff logout from the module. |
| xviii. | The Taxpayer logon again and use the Reference Code to print out the original receipt of the Tax paid. |
| xix. | The taxpayer logout. |

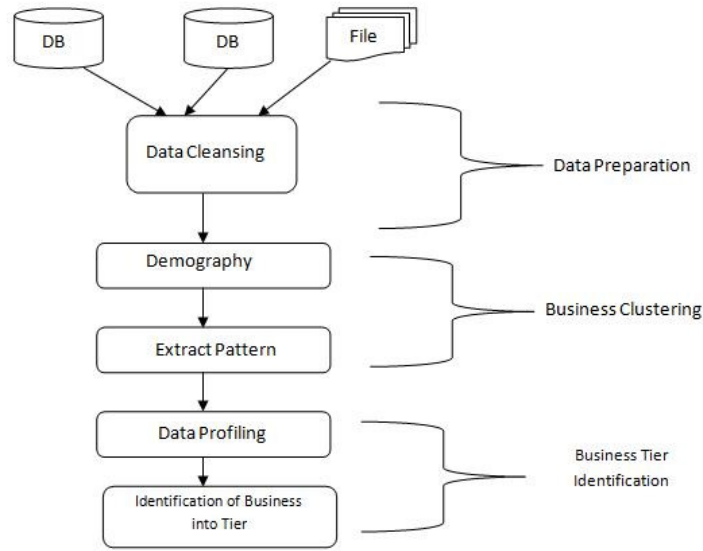

Figure 2 Architectural Design of business profiling using Data collected

The architectural design in Figure 2 depicts the business profiling using data collected. The diagram shows the various stages of using data collected to profile a business into tiers.

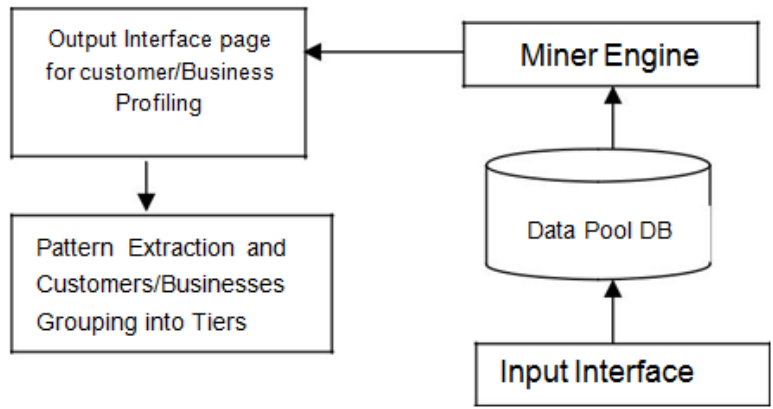

Figure 3 Architectural design of BIR Miner Platform

The diagram in Figure 3 depicts the architecture of the BIR Miner platform. BIR Miner Platform is an active software agent which will extract data set from the data pool, cleanse the data and profile the taxpayer business and group them into Tiers, using the Tax rate guild.

Input Interface Page: This is the interface that captures the data collected into the data pool.

16

Data Pool DB: This is a database where all data collected are stored

Miner Engine: This is the miner platform that will perform the cleansing, the data preparation, clustering and business profiling and grouping into tiers.

Output Interface Page: This is the interface that will display the result of the miner platform.

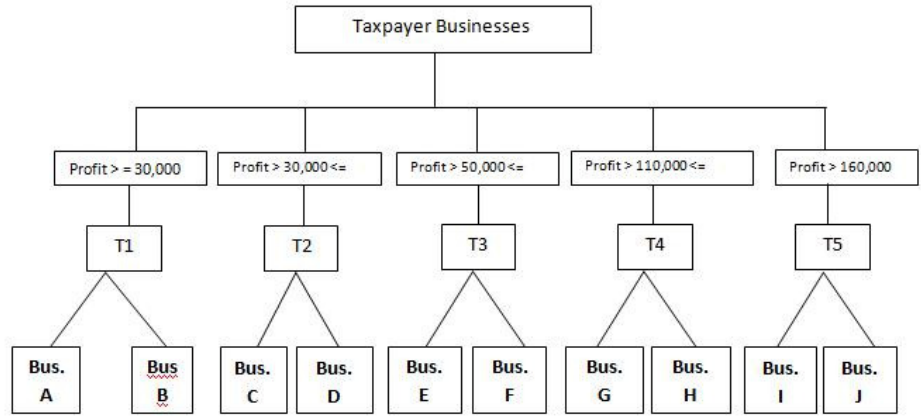

Figure 4 Decision tree classification diagram of taxpayers business

Decision tree builds classification or regression models in the form of a tree structure. It breaks down a dataset into smaller and smaller subset while at the same time a decision tree is incrementally developed. The final result is a tree with decision node and leaf node. Figure 4 depicts the decision tree classification diagram of tax payer business in Benin City, Edo State, Nigeria.

**Taxpayers business:** This is the Root Node of the tree, it represent the business of the taxpayer.

**Profitability expression:** These are the nodes of the tree. It represents the net profit generated from the business of the taxpayers T1 to T5 are the branches of the tree which represent the various tiers in this research work.

**Bus A to Bus J** are the leafs of the tree which represents the group of tax payer businesses that belong to various tiers based on the profit earned.

## 4. FINDINGS AND DISCUSSION

The framework addressed the issues identified with the existing internal revenue collection system of Edo State; other findings are seen as follows from the various architectural modules described below:

### 4.1 BIR (Board of Internet Revenue) Administrators Module

i. The BIR administrator's module can be used in creating the major application users account i.e. BIR staff and bank officials.
ii. The BIR administrator is fully in control of all the functional modules

### 4.2 BIR Staff Module

i. The BIR staff module can be used to capture and store all taxpayers' information to a well-structured database through an input interface.
ii. The BIR staff module can be used to execute a profitability ratio model in order to know the businesses that are making profits or not.
iii. The BIR module of the architecture can be used to properly classify all the taxpayers information stored on the database into their various tiers based on their monthly income/profit according to the clustering technique (decision tree) used in designing and modeling the miner engine
iv. The BIR module can also be used in generating unique TIN (Tax Identification Number) for every taxpayer which will act as a first security check for all logins and tax payment transactions

### 4.3 Tax Payers Module

i. Tax payers' can login to the application after receiving their TIN information through sms alert message or email message as described in the architecture
ii. Each tax payer has a personalized profile in the application to perform different task like request for transaction code to pay tax, payment of tax, receipt for payment and generation / printing of tax clearance.

### 4.4 BIR Bank Module

i. The bank module can perform staff login based on simple authentication scheme
ii. The bank module can capture tax payer's transaction code generated in a text box provided in order to spool out the tax payers details
iii. The module can process the tax payment and commit the transaction to the database
iv. The module can also generate transaction receipt after payment.

### 4.5 BIRGENT Module

i. The intelligent agent module will be able to interact with the database directly in order analyze various transaction activities, detect and deter fraudulent activities like cash suppression and diversion by the generation of unique transaction reference code for tax payers with the well-structured tax payer's details information attached per transaction; which it will also compare with the tax payer's TIN. The agent has well organized reasoning ability.

## 5. RESULTS

This result section uses table 5.1 to explain task to be done by users, action taken by users, and message displayed by the application system.

Table 5.1 Discussion of results

| S/No | Task to be performed by user | Action taken by user | Screen message displayed | Discussions |
|---|---|---|---|---|
| 1. | Login by taxpayer | Correct TIN and correct password enterred | Welcome! | System welcomes and grants taxpayer access to the application based on correct parameters entered |
| 2. | Login by taxpayer | Correct TIN and wrong password or wrong TIN and correct password | Invalid TIN or password…try again | System denies taxpayer access to the application based on incorrect parameters entered |
| 3. | Login by BIR Staff | Correct username and correct password | Welcome! | System welcomes and grants BIR staff access to the application based on correct parameters entered |
| 4. | Login by BIR Staff | Correct username and wrong password or wrong username and correct password | Invalid username or password…try again | System denies BIR staff access to the application based on incorrect parameters entered |
| 5. | Login by Bank staff | Correct username and correct password | Welcome! | System welcomes and grants Bank staff access to the application based on correct parameters entered |
| 6. | Login by Bank staff | Correct username and wrong password or wrong username and correct password | Invalid username or password…try again | System denies Bank staff access to the application based on incorrect parameters entered |
| 7. | Payment of tax by taxpayer | Attempt to change specified tax amount displayed | Amount cannot be altered by taxpayers. | System well structured, tax amount textbox is blocked by the agent against any alteration. Only BIR designated staff in-charge be change that option by performing fresh investigation and review after a period of time as specified by the authority |
| 8. | Mining of taxpayers data by the BIR staff | Spool tax payers' data for mining | Extraction successful! | Useful information are separated from the general data initially captured, TIN generated by the agent and sent to tax payers via sms /email |
| 9. | Clustering of taxpayers into Tiers by the BIR staff | Performing the clustering function without first performing the profitability analysis | Tax payers cannot be clustered into tiers..No records found on earnings or profit margin cluster into tier | The profitability analysis ratio must be processed before the execution of clustering, otherwise there will be no information to |

| 10. | Login by taxpayer | Perform change of password function. Another password entered | Confirmation by re-entering the same password. Password change successful! | Password changed was successful because there was a match in the confirmation. |
| --- | --- | --- | --- | --- |
| 11. | Payment of tax by taxpayer | Correct TIN and Transaction code generated by the agent | Transaction … successful! | The tax payment was successful. Information given were correct |
| 12. | Payment of tax by taxpayer | Correct TIN and wrong transaction code generated by the agent | Fraud Attempt Alert!!! | Intelligent agent detected likely fraudulent attempt. The tax payment cannot be processed any further; transaction aborted. |

## 6. CONCLUSION AND RECOMMENDATIONS

In this paper, we reviewed some known fraud detection techniques in insurance companies, banks, etc. we also examined the tax collection system of Edo State board of internal revenue, and the nature of fraud activities that can be committed. We presented a framework to capture tax payers' data, classify the data into various tiers, detect fraudulent activities like cash suppression and diversion, etc. The framework was designed to have three major components i.e. the BIR Module, Bank Module and the Tax Payers' Module running at the foreground and a software agent called BIRGENT running at the background, which performs a high level of fraud detection and prevention. The framework was designed to have some level of intelligence and reasoning abilities using investigative data mining techniques like Neural Networks, Business classification, decision tree, and so on. We recommend that the architectural framework be developed into a fully functional system using both object oriented programming and agent oriented programming languages, and be deployed by Revenue services at local, state and federal levels to fully combat fraudulent activities in revenue services. The system architectural framework could help in assisting the Government to predict futuristic financial status of the state when fully developed and put into use.